\documentclass[10pt,conference]{IEEEtran}
\IEEEoverridecommandlockouts

\usepackage{adjustbox}
\usepackage{algorithm}
\usepackage{algorithmic}
\usepackage{amsfonts}
\usepackage{amsmath}
\usepackage{amssymb}
\usepackage{array}
\usepackage{colortbl}
\usepackage{diagbox}
\usepackage{float}
\usepackage{graphicx}
\usepackage{hyperref}
\usepackage{inputenc}
\usepackage{lipsum}
\usepackage{makecell}
\usepackage{multirow}
\usepackage{pgf-pie}
\usepackage{pgfplotstable}
\usepackage{pgfplots}
\usepackage{slashbox}
\usepackage{tabularx}
\usepackage{textcomp}
\usepackage{tcolorbox}
\usepackage{tikz}
\usepackage{xcolor}
\usetikzlibrary{patterns, shapes.geometric, arrows, positioning, calc}
\pgfplotsset{compat=1.18}

\begin{document}

\title{Augmenting Large Language Models with Static Code Analysis for Automated Code Quality Improvements}

\author{\IEEEauthorblockN{ Seyed Moein Abtahi}
\IEEEauthorblockA{
\textit{Faculty of Engineering and Applied Science}\\
\textit{Ontario Tech University}\\
seyedmoein.abtahi@ontariotechu.net}
\and
\IEEEauthorblockN{ Akramul Azim}
\IEEEauthorblockA{
\textit{Faculty of Engineering and Applied Science}\\
\textit{Ontario Tech University}\\
akramul.azim@ontariotechu.ca}
}





\maketitle

\begin{abstract} This study examined code issue detection and revision automation by integrating Large Language Models (LLMs) such as OpenAI's GPT-3.5 Turbo and GPT-4o into software development workflows. A static code analysis framework detects issues such as bugs, vulnerabilities, and code smells within a large-scale software project. Detailed information on each issue was extracted and organized to facilitate automated code revision using LLMs. An iterative prompt engineering process is applied to ensure that prompts are structured to produce accurate and organized outputs aligned with the project requirements. Retrieval-augmented generation (RAG) is implemented to enhance the relevance and precision of the revisions, enabling LLM to access and integrate real-time external knowledge. The issue of LLM hallucinations—where the model generates plausible but incorrect outputs—is addressed by a custom-built “Code Comparison App,” which identifies and corrects erroneous changes before applying them to the codebase. Subsequent scans using the static code analysis framework revealed a significant reduction in code issues, demonstrating the effectiveness of combining LLMs, static analysis, and RAG to improve code quality, streamline the software development process, and reduce time and resource expenditure.  \end{abstract}


\section{Introduction}
Natural language processing (NLP) has undergone a significant transformation with the advent of Large Language Models (LLMs) such as OpenAI's GPT series. These models excel in tasks such as text generation, translation, and summarization, and are characterized by their extensive parameters and sophisticated architectures \cite{qin2024large}. Their ability to learn complex language patterns from vast datasets enables the production of human-like text and a nuanced understanding of the context.
Significant advancements have been made in software engineering, particularly in issue detection. Software issues can lead to system failures, financial losses, and security vulnerabilities, necessitating robust detection methods \cite{becker2023programming}. Traditional issue detection, although somewhat effective, often fails to identify the subtle and complex issues inherent in modern software development.
Developers frequently utilize static analysis tools to maintain code quality \cite{yu2024security}.\\
Unlike dynamic analyzers, these tools assess code without executing it, thus providing faster and more predictable analysis. They adhere to predefined rules for style, bug detection, and security, alerting developers to potential issues for prompt rectification. Although manual code reviews are labor-intensive and prone to human error, they play a crucial role in ensuring code quality and maintainability.
Issue detection systems are critical in the software development lifecycle and employ static analysis, dynamic analysis, and machine learning to identify defects early in the development process. Advanced systems offer real-time analysis, seamlessly integrated with development environments to reduce the debugging time and enhance efficiency. Numerous studies have highlighted the effectiveness of these tools for identifying vulnerabilities that may elude manual reviews \cite{9756950}.\\
This study examines the integration of LLMs with issue detection frameworks and, explores methodologies, challenges, and potential improvements in code integrity assurance.
The integration of LLMs into issue detection frameworks presents a promising avenue for enhancing code quality. By leveraging advanced capabilities of LLMs, it is possible to detect more intricate and context-sensitive issues that traditional methods can overlook. This integration could lead to the development of more intelligent and adaptive issue-detection systems capable of learning from past data and continuously improving their detection capabilities. Furthermore, the application of LLMs in this domain could facilitate a more efficient and accurate issue resolution, ultimately leading to more robust and secure software systems. 

In the following sections, a concise review of the related work and background is provided to establish the context for this study. A detailed discussion of the methodologies and the experimental setup employed follows this. Finally, the results are presented and analyzed to provide insights into current practices and identify potential areas for further development.

\section{Background and Related Works}
Large Language Models (LLMs), such as GPT-3.5 Turbo \cite{gpt3} and LLaMA \cite{touvron2023llama}, have demonstrated exceptional capabilities for natural language understanding and complex task generation \cite{ rouzegar2024enhancing}. This section provides an overview of LLMs, focusing on scaling laws, emergent abilities, and key techniques.

Issue detection systems utilize a combination of static and dynamic analyses, along with machine learning, to identify defects early in the development process \cite{9756950}. Advanced systems offer real-time analysis and seamlessly integrate with development environments, thus enhancing efficiency. Research has shown that these systems uncover vulnerabilities \cite{fang2024llm, islam2024llm} that might be overlooked during manual inspections \cite{9756950}. Static analysis tools \cite{panichella2015would}, \cite{davila2021systematic}, in particular, play a crucial role in maintaining code quality by analyzing code without execution, providing faster results than dynamic analyzers \cite{lacombe2024combining}. These tools employ predefined rules to enforce coding standards, detect bugs \cite{srinivasarao2024software}, and identify security issues \cite{tevis2004methods}, subsequently reporting violations for developers to address.

Large language model (LLMs) have become increasingly adept at generating source code in various programming languages, significantly boosting developer productivity. Ensuring the quality of AI-generated code \cite{clark2024quantitative} is crucial, and static analysis tools are instrumental in this regard. Transformer-based LLMs, such as ChatGPT \cite{radford2019language}, can recommend code snippets, although the quality of these recommendations necessitates further evaluation. The readability of code snippets is paramount for their effective use \cite{piantadosi2020does}, with some engines ranking snippets based on their readability scores. Studies suggest that LLMs can enhance code-change tasks by improving the understanding of code modifications \cite{fan2024exploring}.

LLM-specializing techniques, such as domain-specific fine-tuning, enable models to generate precise and context-aware responses for tasks such as code revision or test case generation \cite{ling2023domain, clark2024quantitative}. Prompt engineering, which involves optimizing the prompts to maximize the effectiveness of the generated responses, is pivotal in enhancing model performance \cite{wang2023review, giray2023prompt}. Another key technique is retrieval-augmented generation (RAG), in which a language model is coupled with a retrieval mechanism. This approach enhances the model's ability to generate accurate and contextually relevant responses by accessing external information \cite{chen2024code} before generating the final output. In RAG, the system first retrieves relevant documents \cite{zhao2024retrieval} from external sources, which are then used by the model to produce RAG-bAsed \cite{daneshvar2024exploring} and additional fact-based and reliable answers. This technique improves both the factual accuracy and contextual relevance of LLMs, particularly in tasks that require real-time or domain-specific knowledge \cite{reynolds2024improving}.

Code review is a critical process for ensuring code integrity, although it is often time-consuming \cite{turzo2024makes}. Automation of code reviews using neural LLM has been proposed \cite{rasheed2024ai}, although the current models exhibit limitations. LLMs have the potential to improve code composition, review, and optimization \cite{lu2023llama} by leveraging their extensive knowledge of coding standards and best practices \cite{rasheed2024ai}. Despite this potential, the application of LLMs in code review and optimization remains relatively underexplored. Traditional code review tools frequently lack the depth required for actionable feedback, highlighting the need for LLM-based models to enrich reviews by identifying issues, suggesting optimization, and educating developers.

\begin{figure}[ht]
\centering
\begin{tikzpicture}[node distance=1.5cm, scale=0.8, every node/.style={scale=0.75}]
\tikzstyle{start} = [rectangle, rounded corners, minimum width=3.5cm, minimum height=1.0cm, text centered, draw=black, fill=green!30]
\tikzstyle{process} = [rectangle, minimum width=3.0cm, minimum height=1.0cm, text centered, text width=2.5cm, draw=black, fill=cyan!30]
\tikzstyle{decision} = [diamond, aspect=2, minimum width=3.0cm, minimum height=1.0cm, text centered, draw=black, fill=yellow!50]
\tikzstyle{arrow} = [thick,->,>=stealth]
\tikzstyle{stop} = [rectangle, rounded corners, minimum width=3.5cm, minimum height=1.0cm, text centered, draw=black, fill=red!40]

\node (start) [start] {Project Code};
\node (extract) [process, below of=start] {Issue Analysis and  Extraction};
\node (categorize) [process, below of=extract] {Process Issues and Information};

\node (bugs) [process, below of=categorize, xshift=-4cm] {Bugs};
\node (vuln) [process, below of=categorize] {Vulnerabilities};
\node (smells) [process, below of=categorize, xshift=4cm] {code smells};

\node (prompt_eng) [process, below of=vuln, yshift=-0.5cm] {Prompt Engineering};
\node (rag_decision) [decision, below of=prompt_eng, yshift=-0.5cm] {Use RAG?};

\node (rag_yes) [process, below of=rag_decision, yshift=-0.5cm] {Query Search};
\node (ret_yes) [process, below of=rag_yes] {Retrieve Solutions};
\node (gpt35_yes) [process, below of=ret_yes] {GPT-3.5 Turbo Revision};
\node (resolved_check_yes) [decision, below of=gpt35_yes, yshift=-0.5cm] {Issue Resolved?};
-\node (gpt4_yes) [process, right of=resolved_check_yes, xshift=2.5cm] {GPT-4o Revision};
\node (compare_yes) [process, below of=resolved_check_yes, yshift=-1cm] {Code Comparison};
\node (verify_yes) [process, below of=compare_yes] {Manual Verification};
\node (final_check_yes) [decision, below of=verify_yes, yshift=-0.5cm] {Code Hallucination?};
\node (final_yes) [stop, below of=final_check_yes, yshift=-0.5cm] {Final Revised Code};
\node (manual_yes) [process, right of=final_check_yes, xshift=2.5cm] {Manual Corrections};

\draw [arrow] (start) -- (extract);
\draw [arrow] (extract) -- (categorize);
\draw [arrow] (categorize) -| (bugs);
\draw [arrow] (categorize) -- (vuln);
\draw [arrow] (categorize) -| (smells);
\draw [arrow] (bugs) |- (prompt_eng);
\draw [arrow] (vuln) -- (prompt_eng);
\draw [arrow] (smells) |- (prompt_eng);
\draw [arrow] (prompt_eng) -- (rag_decision);

\draw [arrow] (rag_decision) -- node[anchor=east, font=\small] {Yes} (rag_yes);
\draw [arrow] (rag_yes) -- (ret_yes);
\draw [arrow] (ret_yes) -- (gpt35_yes);
\draw [arrow] (gpt35_yes) -- (resolved_check_yes);
\draw [arrow] (resolved_check_yes) -- node[anchor=east, font=\small] {Yes} (compare_yes);
\draw [arrow] (resolved_check_yes) -- node[anchor=south, font=\small] {No} (gpt4_yes);
\draw [arrow] (gpt4_yes) |- (compare_yes);
\draw [arrow] (compare_yes) -- (verify_yes);
\draw [arrow] (verify_yes) -- (final_check_yes);
\draw [arrow] (final_check_yes) -- node[anchor=east, font=\small] {No} (final_yes);
\draw [arrow] (final_check_yes) -- node[anchor=south, font=\small] {Yes} (manual_yes);
\draw [arrow] (manual_yes) |- (final_yes);

\draw [arrow] (rag_decision) -| node[anchor=south, font=\small] {No} ([xshift=2.3cm]rag_decision.east) |- (gpt35_yes.east);

\end{tikzpicture}
\caption{Automated Code Issue Detection and Revision Pipeline}
\label{pipeline}
\end{figure}
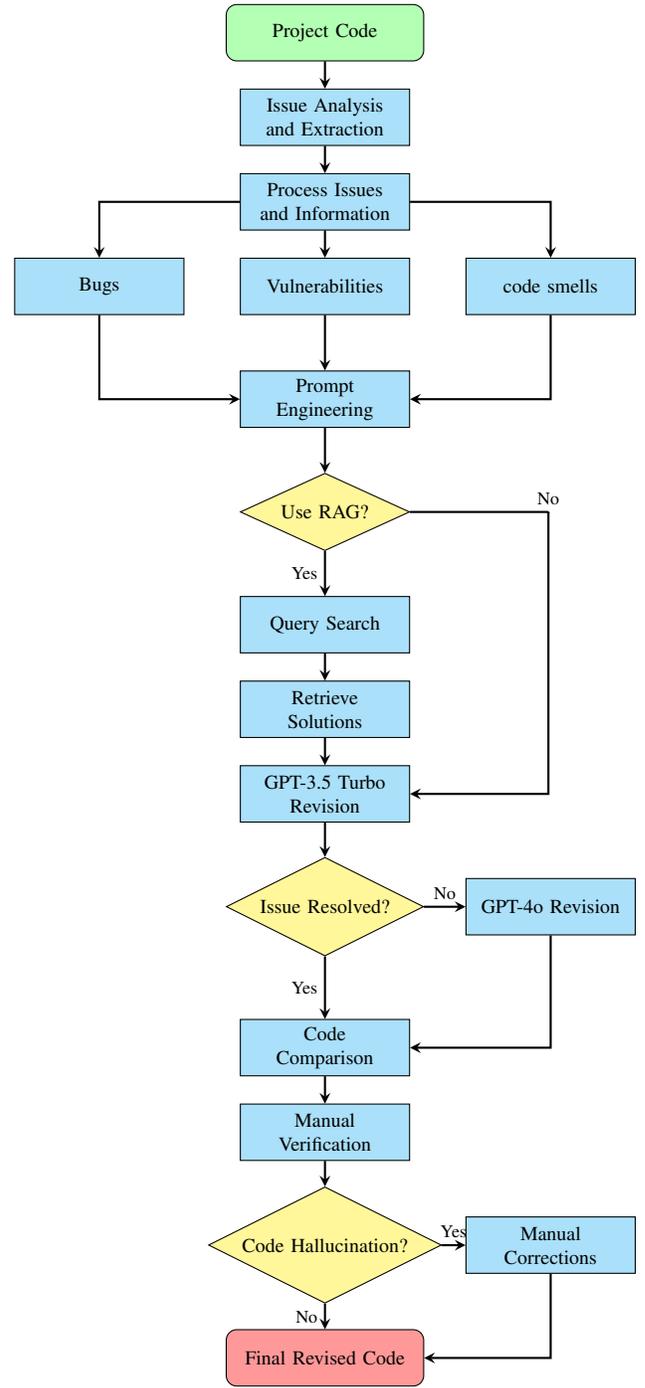

\section{Methodology}

The primary objective of this study is to automate the revision of project code files using advanced large language models (LLMs). Specialized tools for static code analysis and issue detection were employed to identify and extract various types of issues, thereby providing a foundation for automated code revisions.

To automate the process of code revision systematically, an initial analysis of the code dataset was conducted using an issue-detection platform. This platform performs a comprehensive scan of all the code files and, analyzes each line to identify various types of issues. The platform categorizes detected issues into distinct types: \textbf{bugs} (unintentional errors introduced by developers during the coding process), \textbf{code smells} (indicators of potential structural or design problems that suggest the need for refactoring to enhance maintainability), and \textbf{vulnerabilities} (software flaws that present security risks or opportunities for potential exploitation).

Following the identification of issues, detailed information such as file names, line numbers, issue descriptions, and suggested solutions are extracted and documented in data files to support subsequent processing. Properly organizing and processing the extracted data from the code analysis platform is essential for facilitating effective interaction with LLMs.

An overview of this pipeline is shown in Fig. \ref{pipeline}, which begins with a project analysis and concludes with a revised version of the code addressing the identified issues. There are two manual steps in this pipeline. The first step is verification, in which it is assumed that no file contains test cases, necessitating manual verification even for the original files. Based on this approach, if any changes introduced by the contain hallucinations or errors, the next manual phase, correction, is activated. This phase operates in a loop and, requires human intervention to address any problems or issues generated during the LLM revision process.

\subsection{Specializing Large Language Models}

Utilizing LLMs for specific processes or tasks requires specialized techniques to optimize their performance. The first step in this process is prompt engineering, which involves designing and refining prompts that direct the output of a model. Effective prompt engineering can significantly enhance the response quality by providing clear context and specifying the expected format, tone, and content.

After the prompt engineering phase, the next step involves implementing retrieval-augmented generation (RAG). RAG enhances a model’s capabilities by incorporating information from external sources, thereby improving the relevance and depth of generated responses. By retrieving pertinent data from databases, APIs, or knowledge repositories, the model gains access to up-to-date and accurate information, enabling LLM to generate more informed and contextually enriched outputs.

\subsubsection{Prompt Engineering}

Optimizing prompts significantly influences not only the length and accuracy of the responses but also ensures that the format closely aligns with the desired output. To achieve this, a prompt engineering approach was implemented in two steps to enhance the quality of the responses, thereby making them more aligned with the requirements. By carefully structuring prompts, the responses generated by the LLM are rendered more relevant, accurate, and well-organized, effectively addressing the specific needs and expectations of the task.

The initial step involved specifying the programming language of the code file, such as Java or JavaScript. Subsequently, tasks are delineated, focusing on addressing code issues based on details extracted from code analysis platforms (e.g., issue lines, issue types, issue descriptions, and recommended solutions). These tasks were defined as distinct multiple tasks rather than being consolidated into a single paragraph, and a few-shot learning \cite{parnami2022learning} approach was employed.

By providing several examples within the prompt, the LLM model can follow the format of these examples, generating responses that adhere to the desired structure and fulfill the specified requirements. This methodology incorporates sections for detailed instructions and specific tasks, covering three types of issues: bugs, code smells, and vulnerabilities. Each type includes one example of the issue and its corresponding solution, thus enhancing the prompt's capacity to elicit accurate and structured responses from LLM. Each example features a description of the issue, its rationale, correct format, and corresponding corrected code. With this refined version of the prompt, the retrieval-augmented generation \cite{lewis2020retrieval} technique can be applied to further improve response quality.

\begin{algorithm}
\caption{Proposed RAG Technique}
\label{RAG}
\begin{algorithmic}[1]
\REQUIRE Issue description: \texttt{issue\_description}, Issue type: \texttt{issue\_type}
\ENSURE Retrieve a solution for the issue using external sources

\STATE \textbf{Step 1: Formulate Search Query}
    \STATE Extract key terms from \texttt{issue\_description} and \texttt{issue\_type}, and create a targeted search query

\STATE \textbf{Step 2: Retrieve Information}
    \STATE Search external sources (e.g., APIs, documentation, search engines) for possible solutions

\STATE \textbf{Step 3: Filter and Rank Solutions}
    \STATE Filter solutions based on relevance and credibility of the source
    \STATE Rank solutions by factors like source reliability, relevance, and redundancy across multiple sources

\STATE \textbf{Step 4: Augmented Generation}
    \STATE Select the top-ranked solutions and integrate them into the prompt clearly and concisely
    \STATE Add relevant details to strengthen the prompt and improve solution effectiveness based on the retrieved data and other parts of prompt.

\end{algorithmic}
\end{algorithm}
\begin{figure*}[ht]
    \centering
    \includegraphics[width=1.0\linewidth]{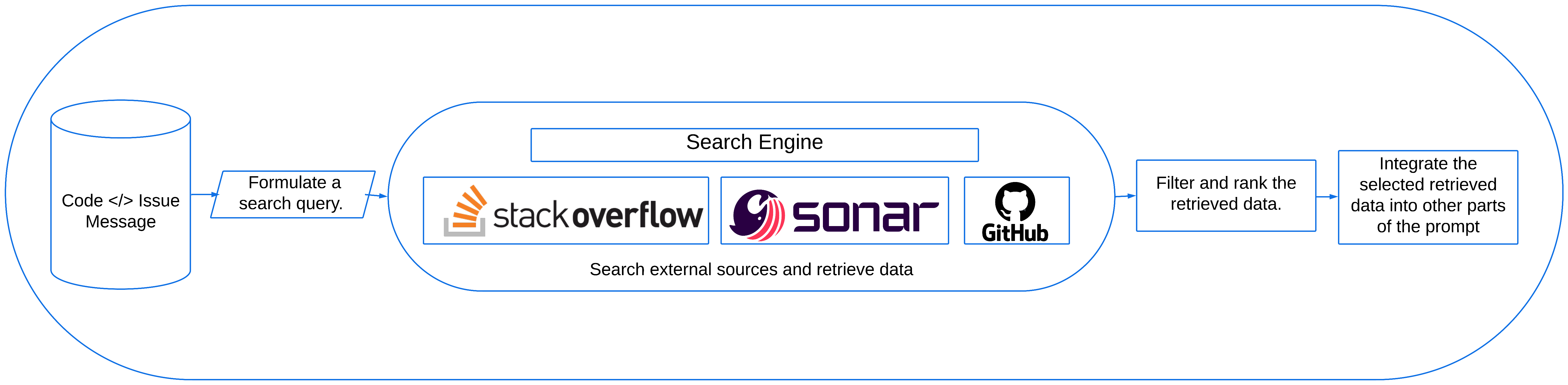}
    \caption{Retrieval-Augmented Generation process}
    \label{fig:RAG1}
\end{figure*}

\subsubsection{Retrieval-Augmented Generation}
The second phase in specializing LLMs involves yielding more coherent responses by employing a retrieval-augmented generation (RAG) approach. This method incorporates an additional search step prior to sending the prompt to the LLM. The retrieval mechanism can vary, producing different responses based on the external source utilized, such as documentation websites, search engines, or APIs. Through this approach, relevant information is retrieved from external datasets and integrated into the prompt before being processed by the LLM API.

A search query, derived from the issue description and relevant details, is employed to extract pertinent information from external sources using the RAG technique. RAG is particularly valuable for producing accurate and explainable responses, as it enables the model to access real-time, up-to-date information \cite{google2024llmspecialization} and solutions for queries that may extend beyond its original training data. The retrieved data, which includes various solution approaches, were synthesized and evaluated. The ranking of the retrieved information and solutions is based on the credibility of the external sources, These ranked solutions are then integrated into the prompt, optimizing it for effective processing by the LLM. In cases where no relevant solutions are retrieved from external sources, the issue is addressed solely by using the LLM’s pre-trained knowledge.

After preparing prompts using the extracted information from the issue analysis platform for each code file, including all relevant code lines, applying prompt engineering techniques, and incorporating data retrieved from external sources via the RAG approach, the next step is to integrate the prompt with LLMs. This integration allows the model to revise the code files based on structured and enriched prompt. The pipeline utilizes the most cost-effective and older LLMs for the first revision. For any remaining issues, a second revision will be conducted using the strongest available LLM to ensure comprehensive and accurate code refinement. Also, the process of using the proposed RAG approach is shown in Alg. \ref{RAG}. and in Fig. \ref{fig:RAG1}.

One of the challenges encountered when using LLMs for code revision is the phenomenon of hallucinations, where the model generates information that appears plausible but is incorrect or not grounded in the provided data. Hallucinations can introduce new issues into the code or provide inaccurate fixes, thereby compromising the reliability of the revisions. To mitigate this, the generated responses (revised code files) from the LLMs were compared to the original code files using a self-developed application based on LLMs measure metrics. This application provides a visual representation of changes and line differences, facilitating the identification and correction of any nonsensical or erroneous modifications introduced by LLMs.

Compared with the original code files, the evaluation of the generated responses was assessed using standard performance metrics such as precision, recall, and F1-score \cite{hu2024unveiling}. These metrics provide valuable insights into the accuracy and effectiveness of the code revisions generated by the LLMs. The overall accuracy of the generated files can be determined based on the values of these metrics, specifically the F1-score. These metrics enable a quantitative assessment of how effectively LLMs address the detected issues and, clearly understand the model's performance in the code revision process.

If the Code Comparison App confirms that the generated files are free from hallucinations, they can be used as suggested files to improve the original code or even as replacements for the original code. However, if hallucinations are detected, the generated files undergo human review and correction as the final step. Once human revisions were completed, the files are ready for use.
As illustrated in Fig. \ref{pipeline}, the overall methodology aims to streamline software development cycles by enhancing issue detection, facilitating timely optimization, and automating extensive processes. The ultimate goal is to significantly reduce the time required to revise code files efficiently, thereby improving software quality and maintainability. By leveraging the LLMs, the process transforms the original files into reliable final revisions. The the following section presents and discusses, the implementation of the pipeline.

\begin{table*}[ht]
    \centering
    \caption{Example Report Format for Code Quality Issues Detected by SonarQube}
    \resizebox{\textwidth}{!}{%
        \begin{tabular}{|c|c|c|c|c|}
            \hline
            \textbf{File\_Location} & \textbf{File\_Name} & \textbf{Line} & \textbf{Message} & \textbf{Type} \\
            \hline
            \makecell{EIS\textbackslash{}client\textbackslash{}src\\\textbackslash{}components\textbackslash{}BaseMap\\\textbackslash{}vehicleMarkers.jsx} &
            Revised.vehicleMarkers.jsx & 409 & Visible, non-interactive elements with   click handlers must have at least one keyboard listener. & BUG \\ \hline
            \makecell{EIS\textbackslash{}deploy\textbackslash{}helm\textbackslash{}eis\\\textbackslash{}charts\textbackslash{}eis-backend\textbackslash{}templates\\\textbackslash{}deployment.yaml} & Revised.deployment.yaml & 16 & Set automountServiceAccountToken to false for   this specification of kind Deployment. & VULNERABILITY \\ \hline
            \makecell{EIS\textbackslash{}client\textbackslash{}tests\\\textbackslash{}endtoend\textbackslash{}settings\textbackslash{}rsc.js} & Revised.rsc.js & 361 & Remove this commented out code. & CODE\_SMELL \\ \hline
            
        \end{tabular}
    }
    \label{fig:csv}
\end{table*}

\section{Experimental Setting}


\subsection{Experimental Design and Setup}
Based on the proposed methodology outlined in Fig. \ref{pipeline}, the selection of appropriate tools and datasets was crucial for the experiment. SonarQube was selected as the platform for analyzing the dataset and inspecting the quality of the code files, as it effectively identifies and detects issues within the code. The dataset used in this experiment comprised various code files in different programming languages, allowing for the measurement of the effectiveness of large language models in this context. The LLMs utilized in this experiment were sourced from OpenAI\cite{openai}.

\subsubsection{SonarQube}
SonarQube, an open-source platform developed by SonarSource, enables continuous code-quality inspection through static code analysis \cite{sonarqube}. It categorizes the identified issues into three main types: bugs, code smells, and security vulnerabilities, applicable across various programming languages, including JavaScript. SonarQube also implements quality gates as checkpoints to enforce the coding standards and quality metrics. 

An initial scan of the original dataset of the project classified code issues into three categories: bugs, code smells, and vulnerabilities. Three data files are generated for each category. The data files extracted from SonarQube include file names, locations, issue lines, SonarQube's suggested solutions, and issue types. This meticulous documentation was aimed at enhancing response accuracy within the LLM.

The data extracted from SonarQube was used to accurately identify specific code issues within particular files (documented as ``file\_name" in the data file), precise lines (documented as ``line" in the data file), and detailed descriptions (documented as ``message" in the data file). Additionally, the type of issue (recorded as ``type" in the data file) and the file path (recorded as ``file\_location" in the data file) within the project were included. This information facilitates the recreation of files based on these paths, as listed in Table \ref{fig:csv}.

It is important to note that these data files were customized and explicitly modified for experimental purposes, and did not represent the raw data directly from SonarQube. This dataset forms the basis for generating prompts for communication with the OpenAI API.

\subsubsection{Dataset}
Project code files were employed as the primary source for this experiment, known as the ``EIS" dataset from Team Eagle \cite{team-eagle}. Utilizing all available files from this project allowed for comprehensive testing across various programming languages and file types, including JavaScript, Python, Java, and Docker configuration files. This approach ensured the thorough exploration and analysis of diverse code functionalities and interactions within the project environment.

\subsubsection{Large Language Models and API}
The LLMs selected for this experiment were GPT-3.5 Turbo\cite{gpt3} and GPT-4o\cite{gpt4}. The GPT-3.5 Turbo is a general-purpose, cost-effective model and one of the older options in the OpenAI lineup. In contrast, GPT-4o represents the most recent and powerful model available, offering advanced capabilities that exceed those of the other models. The rationale behind selecting these two models is to leverage GPT-3.5 Turbo to revise the majority of files under cost-effective conditions. For unresolved issues, GPT-4o was employed to ensure that more complex problems were addressed with greater computational strength.

In large-scale experiments utilizing LLMs, reliance on the graphical user interface (GUI) of the models is inadvisable; instead, an API should be employed. The challenge with the API is the lack of cached data and history of requests and responses. To mitigate costs, it is essential to formulate optimal prompts that effectively elicit desired responses. This approach involves consolidating all the issues into a single file, transmitting the complete original code, and receiving the revised file in its original format without additional comments in a single API request.

\subsubsection{Prompt Engineering }
Prompt engineering is an iterative process that requires extensive testing to develop prompts that can elicit accurate responses from LLMs. In this study, more than 20 iterations were performed to refine the final prompt. 
Fig. \ref{fig:prompt_template} presents an overview of the prompt engineering process.

The final version of the prompt includes all the relevant code lines, identifies specific code issues, and provides detailed descriptions and solutions while addressing multiple issues within a file. It employs few-shot learning and, features three revision examples for each issue type and output format. Tasks were divided into simpler sub-tasks to improve clarity. Data extracted from SonarQube, presented in Table \ref{fig:csv} inform the final prompt format, which is used in the \ref{Experiment Implementation} section and is outlined in detail in Algorithm \ref{FixCodeIssuesWithRAG}. The predefined system and user roles were integrated into the API to optimize the results.

\begin{figure}[ht]
    \centering
    
    \begin{tcolorbox}[colframe=black!75!white, colback=white!95!gray, title=Iterative Prompt Engineering Process]
    
    \textbf{Stage 1: The Initial Prompt} \\
    Provides the full code and describes issues. Ask for revisions and a message confirming the file generation.
    \begin{itemize}
        \item \textbf{Input:} Full code file
        \item \textbf{Issues:} Issues  description
    \end{itemize}
    \vspace{0.2cm}
    \vspace{0.3cm}
    \hrule
    \vspace{0.3cm}

    \textbf{Stage 2: Structured Issue Identification} \\
Organize code issues by line and type with brief descriptions. Specify the language, request step-by-step revisions, confirm after each update, and generate a revised file.
    \begin{itemize}
        \item \textbf{Issue 1.BUG:} Line 116, Visible, non interactive elements with click handlers must have at least one keyboard listener.

        \item \textbf{...}
    \end{itemize}
    \vspace{0.2cm}
    
    \vspace{0.3cm}
    \hrule
    \vspace{0.3cm}
    
    \textbf{Stage 3: Few-shot Learning with Examples} \\
    Further refine by including examples of correct fixes for each type of issue.
    \begin{itemize}
        \item \textbf{Example 1 (BUG):} \\
        Original: \texttt{onClick=\{handleClick\}} \\
        Corrected: \texttt{<div onClick=\{handleClick\} onKeyDown=\{handleKeyDown\} tabIndex="0">...</div>}
        \item \textbf{...} \\
    \end{itemize}
    \textbf{Evaluation:} Comprehensive prompt with structure and examples
    
    \end{tcolorbox}
    
    \caption{Illustration of the iterative prompt engineering process}
    \label{fig:prompt_template}
\end{figure}

\subsubsection{Retrieval-Augmented Generation}

The retrieval-augmented generation (RAG) technique enhances the quality of responses generated by LLMs by leveraging external sources such as Stack Overflow, GitHub, the SonarQube Community, and Google search platforms. These sources are highly regarded by the software development community for their extensive discussions on coding, debugging, and software engineering, offering valuable insights and technical data.

In this technique, queries are constructed based on the descriptions of identified issues and are used to search for external sources of solutions. Four external sources were utilized in this experiment: GitHub Community API, SonarQube Community, and Stack Overflow. Additionally, Google’s search engine was used to search for broader web of issue-related information. The first three sources, GitHub, SonarQube, and Stack Overflow, were prioritized because they are dedicated programming libraries and documentation repositories. If no results were found in these primary sources, the system then utilized the Google Search API to retrieve information from the web, focusing on the top three ranked pages, where the likelihood of finding relevant answers was higher. The retrieved information was further validated by analyzing the source URLs and ranking the data based on the credibility of the source, as determined by Google’s ranking system. 

Solutions are ranked according to the credibility of the source: GitHub was ranked highest, followed by SonarQube Community, Stack Overflow, and other sources accessed through Google search. The curated data were then integrated into the LLM prompt to provide the model with accurate and contextually relevant technical information. This approach significantly enhances the quality and precision of a model's responses. If no relevant solutions were identified through this Retrieval-Augmented Generation (RAG) process, the issue was revised based solely on the LLM's pre-trained knowledge, ensuring the system's ability to revise the code effectively even in the absence of external information. Furthermore, the RAG process exclusively utilizes issue descriptions to find solutions, ensuring that no information leakage occurs during the implementation of this technique.

\subsubsection{Evaluation Metrics}

A ``code comparison" application was developed to assess the accuracy of the files generated by LLMs. This application facilitates a side-by-side comparison between the original and revised files, enabling the identification of the differences between each file and its updated version. This approach simplifies the evaluation of revision accuracy by highlighting the variations between the original and revised code lines, thereby determining the viability of using LLMs for code file revisions. In addition, the application aids in detecting code hallucinations by clearly marking discrepancies, allowing quicker edits and revisions as the final step in the pipeline. The metrics used in the application, including precision, recall, and F1-score, collectively assessed the accuracy of the generated files. These metrics were used to assess the percentage of changes in the revised version of each file compared with the original version with issues. Although designed for classification problems, they also help to identify which files have more issues and a greater number of changes. A metric value close to 100 percent indicates fewer changes, where lower values indicate more extensive modifications.

\begin{algorithm}
\caption{Procedure for Fixing Code Issues Using LLM API with Prompt Engineering and RAG}
\label{FixCodeIssuesWithRAG}
\begin{algorithmic}[1]
\REQUIRE File location: \texttt{file\_location}, Original code: \texttt{original\_code}, Issue lines: \texttt{issue\_lines}, Issue messages: \texttt{issue\_messages}, Issue types: \texttt{issue\_types}
\ENSURE Corrected code with all issues resolved
\STATE \textbf{Task:} ``Specify the programming language and address code issues based on specific details."
\STATE \textbf{Example 1:} ``Fix a Bug in JSX on Line 409: \texttt{<div onClick=\{handleClick\}>...</div>}'' 
\newline 
\textbf{Corrected:} \texttt{<div onClick=\{handleClick\} onKeyDown=\{handleKeyDown\} tabIndex="0">...</div>}

\STATE \textbf{Ensure:} ``Maintain the original format of the corrected code. Follow provided examples to resolve the issues."

\FOR{each issue: \texttt{issue\_line}, \texttt{issue\_message}, \texttt{issue\_type}}
    \STATE \textbf{Issue Description:} ``The detected \texttt{issue\_type} issue by SonarQube is on line: \texttt{issue\_line}. Description: \texttt{issue\_message}"
    
    \IF{RAG is applicable}
        \STATE \textbf{Retrieve Context:} Use RAG techniques proposed in Alg. \ref{RAG} to retrieve and integrate relevant solutions from external sources for \texttt{issue\_type}.
        \STATE \textbf{Example Guidance:} Follow retrieved examples to formulate final response.
    \ENDIF
\ENDFOR
\STATE \textbf{Final Instructions:} ``If similar issues are present on other lines, please fix those as well based on retrieved context."
\STATE \textbf{Final Display:} ``Here is the corrected code based on the aforementioned issues and examples. Only corrected code sections will be provided."
\end{algorithmic}
\end{algorithm}

\subsection{Experiment Implementation}
\label{Experiment Implementation}
After preparing all the necessary tools, setting up the required files, and finalizing the prompt, the entire ``EIS" folder was re-scanned using SonarQube to extract all issues. The initial scan identified 234 bugs, 61 vulnerability issues, and 7,304 code smells. The project was then divided into four parts based on the types of issues: the first part focused on preparing the prompt for bug-related issues; the second part addressed code smell issues; the third part dealt with vulnerability issues, and the final part included all types of issues to enable a comparative analysis of the results. Each part was conducted in two phases: the first phase utilized the GPT-3.5 Turbo model and the second phase employed the GPT-4o model for the remaining issues.

Before proceeding with the remaining steps of the proposed methodology, the experiment was divided into two parts, following the extraction of issues using SonarQube. The first part involved utilizing RAG for code revision with LLMs, whereas the second part relied solely on prompt engineering without RAG. By separating these two approaches, a comparative analysis of their results was conducted, enabling the evaluation of the advantages associated with each method. The outcomes of this comparison are presented in Table \ref{tab:results}.

\subsubsection{Bug Analysis}

The ``EIS" folder was initially scanned, which revealed 234 bugs. Following the first revision using GPT-3.5 Turbo, a new directory, ``EIS2," was created to house the revised files. A subsequent SonarQube scan of the ``EIS2" folder identified and cataloged 117 remaining bugs, that were documented in a data file. In the second revision, all remaining bugs were further refined using GPT-4o. A follow-up SonarQube scan confirmed that the ``EIS2" folder was free of bugs, demonstrating a significant reduction in issues and ensuring the enhanced reliability of the software.

\subsubsection{Vulnerability Analysis}

The initial scan of the ``EIS" folder identified 61 vulnerability issues. After the first revision using GPT-3.5 Turbo, a new directory, ``EIS3," was created to store the revised files. A SonarQube scan of the ``EIS3" folder found that only two vulnerabilities remained, which were documented in the data file. A second revision using GPT-4o was performed to address the remaining vulnerability issues. A subsequent SonarQube scan confirmed that all vulnerabilities were successfully resolved, ensuring the security and integrity of the ``EIS3" folder, thereby bolstering the system's resilience against potential security threats.

\subsubsection{Code smell Analysis}

An initial analysis of the ``EIS" folder revealed 7,304 code smell issues. The first revision, performed using GPT-3.5 Turbo, resulted in the revised files being stored in a new directory named ``EIS4." A subsequent SonarQube scan of the ``EIS4" folder, identified 3,586 remaining code smell issues that were documented in the table format file. To address these issues, a second revision was conducted using the GPT-4o. A final SonarQube scan confirmed that 1,367 code smell issues remained, significantly improving the quality and maintainability of the ``EIS4" folder and thereby enhancing the readability and maintainability of the codebase.

\subsubsection{Comprehensive Analysis}
In the final phase of the test, before utilizing an LLM to revise all categories of issues within the ``EIS" folder, a comprehensive scan was conducted using SonarQube. The initial scan identified 7,599 issues. Prompts were meticulously prepared for each file based on these issues. For the initial revision, all files were submitted to the API using the GPT-3.5 Turbo model, resulting in the creation of a revised folder labeled ``EIS.Revised". 

A follow-up scan of the ``EIS.Revised" folder with SonarQube revealed a reduction in issues to 4,340. To further address the remaining issues, prompts were prepared for each file, encompassing all identified problems. In this stage, the GPT-4o model was used for the revision. Upon re-scanning the ``EIS.Revised" folder with SonarQube, the number of unresolved issues was further reduced to 1,058.

\subsection{Evaluation}

Upon completion of file revisions by LLMs, the next critical step is to verify the accuracy and validity of the generated files to ensure their usability. This was achieved using a custom-developed application with two primary functions. First, the application compares all the generated files with the original files, highlighting the differences between the code versions to facilitate the identification of changes. Second, it calculates and displays key evaluation metrics such as precision, recall, and F1-score for each file, allowing for individual file assessment. This enables faster final revisions by a human reviewer, thereby significantly reducing the time required to inspect differences and mitigate potential LLM hallucinations. The average values for these evaluation metrics are presented in Table \ref{tab:results}.
Thorough validation is critical, as invalid files cannot be source files in production environments. The automated comparison and evaluation process enhances the efficiency and effectiveness of the verification phase, ensuring that the LLM-generated code meets the required standards and can be reliably used as a software project files.
The next section contains all the results of this experiment.

\section{Results and Analysis}

To ensure fairness and validate that GPT-4o can also address any issue resolved by GPT-3.5 Turbo, a separate validation experiment was conducted, and the results are presented in Table \ref{tab:result}. The findings confirm that GPT-4o, as a more advanced model, resolves all issues handled by GPT-3.5 Turbo and additional issues that GPT-3.5 Turbo could not address. Given the higher cost of GPT-4o, a cost-efficient approach was adopted: GPT-3.5 Turbo was used for the initial revisions, with GPT-4o reserved for unresolved issues. This strategy ensures effective results while minimizing costs.

\begin{table}[ht]
\footnotesize
\renewcommand{\arraystretch}{1.4}
\caption{Comparative Analysis of GPT-3.5 Turbo and GPT-4o for Issue Resolution: Revision Coverage and Cost Efficiency}
\label{tab:result}
\centering
\resizebox{\columnwidth}{!}{
\begin{tabular}{|>{\centering\arraybackslash}m{4.5cm}|c|c|c|}
\hline
\multirow{2}{*}{\textbf{Metrics}} & \multicolumn{3}{c|}{\textbf{Issues Category}} \\
\cline{2-4}
 & \textbf{Bugs} & \textbf{Vulnerabilities} & \textbf{Code Smell} \\ \hline
\textbf{Total Issues in Sample Dataset} & 30 & 30 & 30 \\ \hline
\multicolumn{4}{|c|}{\cellcolor{gray!20}\textbf{Revision Coverage}} \\ \hline
GPT-3.5 Turbo & \textbf{26 / 30} & \textbf{28 / 30} & \textbf{29 / 30} \\ \hline
GPT-4o & \textbf{30 / 30} & \textbf{30 / 30} & \textbf{30 / 30} \\ \hline
\multicolumn{4}{|c|}{\cellcolor{gray!20}\textbf{Cost Analysis (USD)}} \\ \hline
GPT-3.5 Turbo Rev. Cost & \$0.41 (26 Revisions) & \$0.16 (28 Revisions) & \$0.08 (29 Revisions) \\ \hline
\begin{tabular}[c]{@{}c@{}}GPT-4o Rev. Cost for Exact\\Files Revised by GPT-3.5 Turbo\end{tabular} 
& \$0.71 (26 Revisions) & \$1.82 (28 Revisions) & \$0.25 (29 Revisions) \\ \hline
GPT-4o Rev. Cost & \$0.82 (30 Revisions) & \$2.04 (30 Revisions) & \$0.26 (30 Revisions) \\ \hline
\multicolumn{4}{|c|}{\cellcolor{gray!20}\textbf{GPT Models Average Cost per Issue (USD)}} \\ \hline
GPT-3.5 Turbo Average Rev. Cost Per Issue & \$0.015 & \$0.006 & \$0.003 \\ \hline
GPT-4o Average Rev. Cost Per Issue & \$0.027 & \$0.065 & \$0.008 \\ \hline
\end{tabular}
}
\end{table}

The performance analysis of GPT-3.5 Turbo and GPT-4o in revising code issues across various categories yielded valuable insights into their capabilities and efficiencies. A detailed summary of the results is presented in Table \ref{tab:results}.

\begin{table*}[htbp]
\centering
\caption{GPT Model Revision Analysis: The table presents a comprehensive analysis of revisions performed by GPT-3.5 Turbo and GPT-4o on different categories of issues, including bugs, vulnerabilities, and code smells. It highlights the number of issues identified, files revised, revision success rates, and associated costs.}
\label{tab:results}
\scalebox{1.12}{ 
\begin{tabular}{|l|c|c|c|c|}
\hline
\diagbox{\textbf{Metric}}{\textbf{Issue Category}} &
  \textbf{EIS2 (Bugs)} &
  \textbf{EIS3 (Vuln.)} &
  \textbf{EIS4 (C\_Smell)} &
  \textbf{EIS.Rev (Comprehensive)} \\ \hline
Total \# Issues in EIS                     & 234     & 61      & 7,304     & 7,599     \\ \hline
\# Issues Rev. by GPT-3.5 Trubo Only       & 117     & 59      & 3,718     & 3,259     \\ \hline
\# Remainig Issues Rev. by GPT-4o Only            & 117     & 2       & 2,219     & 2,186     \\ \hline
\# Issues Rev. ( GPT-3.5 Turbo + GPT-4o )    & 234     & 61      & 5,937     & 5,445     \\ \hline
GPT-3.5 Turbo Success Rate & 50\%    & 96.7\%  & 50.9\%   & 42.8\%   \\ \hline
GPT-4o Success Rate        & 100\%   & 100\%   & 61.8\%   & 50.36\%  \\ \hline
Total GPT Models Success Rate & \textbf{100\%} & \textbf{100\%} & \textbf{81.2\%} & 71.6\% \\ \hline
Ave. Precision in Rev. Files        & 100\%   & 100\%   & 100\%   & 100\%  \\ \hline
Ave. Recall  in Rev. Files       & 97\%   & 98.8\%   & 92.67\%   & 90.06\%  \\ \hline
Ave. F1-Score in Rev. Files       & 98.4\%   & 99.25\%   & 96.05\%   & 94.24\%  \\ \hline
GPT-3.5 Turbo Rev. Cost (USD)   & \$1.68 & \$0.25 & \$8.13 & \$8.05 \\ \hline
GPT-4o Rev. Cost (USD)        & \$3.08 & \$0.13 & \$18.69 & \$17.86 \\ \hline
Total Rev. Cost (USD)       & \$4.76 & \$0.38 & \$26.82 & \$25.91 \\ \hline
\end{tabular}
}
\end{table*}

A total of 7,599 issues were analyzed in this study and categorized into three groups: bugs (EIS2), vulnerabilities (EIS3), and code smells (EIS4). The distribution was highly imbalanced, with bugs accounting for 234 issues, vulnerabilities for 61, and code smells for 7,304 issues. Both GPT models demonstrated complete coverage in addressing bugs and vulnerabilities, successfully revising all 234 bug-related files and 61 vulnerability-related files. For code smells, GPT-3.5 Turbo revised 3,718 files, whereas GPT-4o addressed 2,219 files, indicating a more selective approach by GPT-4o in handling this category.

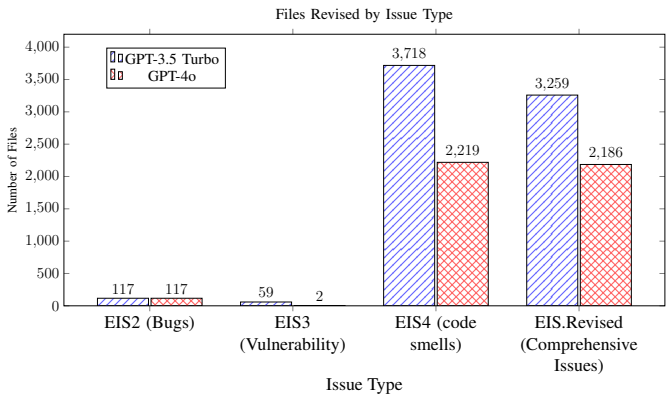
\begin{figure}[ht]
    \centering
    \resizebox{1.0\columnwidth}{!}{ 
    \begin{tikzpicture}
    \begin{axis}[
        ybar,
        bar width=40pt, 
        width=1.0\textwidth,
        height=0.5\textwidth,
        legend style={at={(0.17,0.95)}, anchor=north, font=\large},
        ylabel={Number of Files},
        ylabel style={font=\large},
        xlabel={Issue Type},
        xlabel style={at={(0.5,-0.15)}, font=\Large},
        symbolic x coords={EIS2 (Bugs), EIS3 (Vulnerability), EIS4 (code smells), EIS.Revised (Comprehensive Issues)},
        xtick=data,
        nodes near coords,
        nodes near coords align={vertical},
        every node near coord/.append style={font=\large}, 
        ymin=0, ymax=4200,
        title={Files Revised by Issue Type},
        title style={font=\large},
        tick label style={font=\large},
        xlabel style={at={(0.5,-0.25)}, font=\Large},
        x tick label style={text width=3.5cm, align=center, font=\Large}, 
        enlarge x limits=0.2, 
        ylabel near ticks, 
        y tick label style={/pgf/number format/.cd, fixed, fixed zerofill, precision=0, /tikz/.cd}, 
    ]
    \addplot[pattern=north east lines, pattern color=blue!60] coordinates {(EIS2 (Bugs), 117) (EIS3 (Vulnerability), 59) (EIS4 (code smells), 3718) (EIS.Revised (Comprehensive Issues), 3259)};
    \addplot[pattern=crosshatch, pattern color=red!60] coordinates {(EIS2 (Bugs), 117) (EIS3 (Vulnerability), 2) (EIS4 (code smells), 2219) (EIS.Revised (Comprehensive Issues), 2186)};
    \legend{GPT-3.5 Turbo, GPT-4o}
    \end{axis}
    \end{tikzpicture}}
    \caption{ Files Revised by Issue Type for GPT-3.5 Turbo and GPT-4o} 
    \label{fig:files_revised}
\end{figure}

In addition to evaluating individual issue types, a comprehensive test was conducted to assess the performance of the models when processing all issues within a file, regardless of type, in a single prompt (EIS.Revised). This approach aimed to compare the efficiency of handling mixed-issue prompts against individual issue-type processing and to examine the models' ability to resolve diverse issues collectively. Through this method, the models resolved a total of 5,445 issues, with GPT-3.5 Turbo addressing 3,259 and GPT-4o resolving 2,186. The results are visualized in Fig. \ref{fig:files_revised}, which depicts the number of files revised by each model. In some instances, GPT-4o revised fewer files than GPT-3.5 Turbo, as GPT-3.5 Turbo effectively resolved the majority of issues, leaving only the remaining unresolved issues for GPT-4o to address.

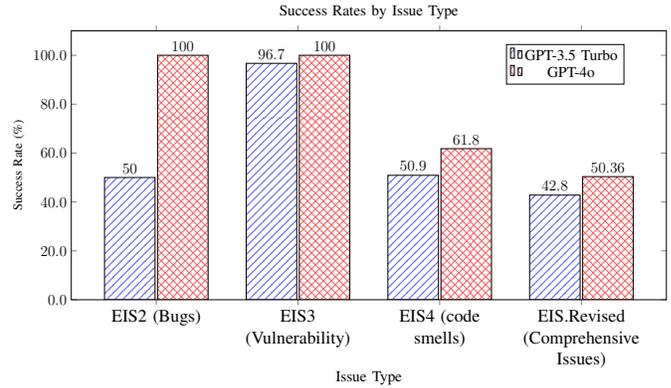
\begin{figure}[ht]
    \centering
    \resizebox{1.0\columnwidth}{!}{ 
    \begin{tikzpicture}
    \begin{axis}[
        ybar,
        bar width=40pt, 
        width=1.0\textwidth,
        height=0.5\textwidth,
        legend style={at={(0.83,0.95)}, anchor=north, font=\large},
        ylabel={Success Rate (\%)},
        ylabel style={font=\large},
        xlabel={Issue Type},
        xlabel style={at={(0.5,-0.25)}, font=\Large},
        symbolic x coords={EIS2 (Bugs), EIS3 (Vulnerability), EIS4 (code smells), EIS.Revised (Comprehensive Issues)},
        xtick=data,
        nodes near coords,
        nodes near coords align={vertical},
        every node near coord/.append style={font=\large}, 
        ymin=0, ymax=110,
        title={Success Rates by Issue Type},
        title style={font=\large},
        tick label style={font=\large},
        label style={font=\large},
        x tick label style={text width=3.5cm, align=center, font=\Large}, 
        enlarge x limits=0.2, 
        ylabel near ticks,
        y tick label style={/pgf/number format/.cd, fixed, fixed zerofill, precision=1, /tikz/.cd}, 
    ]
    \addplot[pattern=north east lines, pattern color=blue!60] coordinates {(EIS2 (Bugs), 50) (EIS3 (Vulnerability), 96.7) (EIS4 (code smells), 50.9) (EIS.Revised (Comprehensive Issues), 42.8)};
    \addplot[pattern=crosshatch, pattern color=red!60] coordinates {(EIS2 (Bugs), 100) (EIS3 (Vulnerability), 100) (EIS4 (code smells), 61.8) (EIS.Revised (Comprehensive Issues), 50.36)};
    \legend{GPT-3.5 Turbo, GPT-4o}
    \end{axis}
    \end{tikzpicture}
    } 
    \caption{Success Rates by Issue Type for GPT-3.5 Turbo and GPT-4o}
    \label{fig:success_rates}
\end{figure}

The success rates of the GPT models varied across issue types. For bugs (EIS2), a 50\% success rate was achieved by GPT-3.5 Turbo, while GPT-4o demonstrated a perfect 100\% success rate. Both models address vulnerabilities (EIS3), with GPT-3.5 Turbo achieving a 96.7\% success rate and GPT-4o reaching 100\%. For code smells (EIS4), success rates of 50.9\% and 61.8\% were achieved using the GPT-3.5 Turbo and GPT-4o, respectively. GPT-4o consistently outperformed GPT-3.5 Turbo across all issue types. The combined efforts of both models resulted in impressive total success rates of 100\% for bugs and vulnerabilities, and 81.2\% for code smells. A comprehensive approach (EIS.Revised) was employed to encompass all issue types. This approach revealed 7,599 issues across all categories. The GPT-3.5 Turbo revised files with a 42.8\% success rate, whereas GPT-4o revised with a 50.36\% success rate. The combined effort resulted in an overall success rate of 71.6\%. More details can be found in Fig. \ref{fig:success_rates}.

\begin{figure}[ht]
    \centering
    \resizebox{1.0\columnwidth}{!}{ 
    \begin{tikzpicture}
    \begin{axis}[
        ybar,
        bar width=50pt, 
        width=1.0\textwidth,
        height=0.5\textwidth,
        ylabel={Accuracy (F1-Score) Rate (\%)},
        ylabel style={font=\large},
        xlabel={Issue Type},
        xlabel style={at={(0.5,-0.25)}, font=\Large},
        symbolic x coords={EIS2 (Bugs), EIS3 (Vulnerability), EIS4 (code smells), EIS.Rev (Comprehensive Issues)},
        xtick=data,
        nodes near coords,
        nodes near coords align={vertical},
        every node near coord/.append style={font=\large}, 
        ymin=0, ymax=110,
        title={Generated File Accuracy by Issue Type},
        title style={font=\large},
        tick label style={font=\large},
        label style={font=\large},
        x tick label style={text width=3.5cm, align=center, font=\Large}, 
        enlarge x limits=0.2, 
        ylabel near ticks, 
        y tick label style={/pgf/number format/.cd, fixed, fixed zerofill, precision=1, /tikz/.cd}, 
    ]
    \addplot[pattern=north east lines, pattern color=blue!60] coordinates {(EIS2 (Bugs), 98.4) (EIS3 (Vulnerability), 99.25) (EIS4 (code smells), 96.05) (EIS.Rev (Comprehensive Issues), 94.24)};
    \end{axis}
    \end{tikzpicture}}
    \caption{LLM Generated File Accuracy (F1-Score) by Issue Type}
    \label{fig:accuracy_rates}
\end{figure}
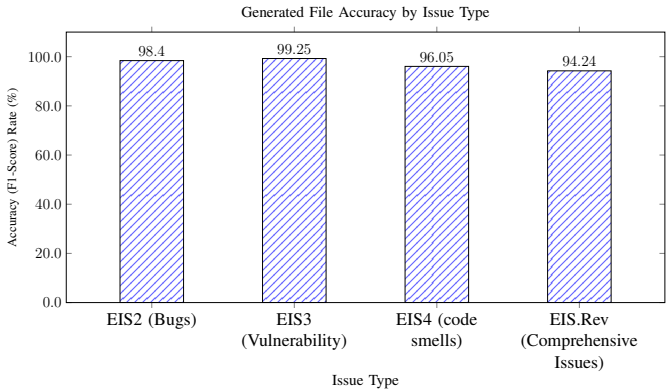

A significant finding of this study is the superiority of the divided approach over the comprehensive approach in terms of accuracy and success rates. Variations in the accuracy of the generated files were noted across issue types and approaches, as shown in Fig. \ref{fig:accuracy_rates}. For the divided approach, where issues were addressed by type, high accuracy rates were observed: 99.25\% for vulnerabilities (EIS3), for bugs (EIS2) 98.4\%, and 96.05\% for code smells (EIS4). These results indicate a high level of reliability in the revisions made by the GPT models when issues are categorized and addressed separately. In contrast, when all types of issues were combined in a single prompt (EIS.Revised), an accuracy rate of 94.24\% was achieved. 

As evidenced by the data in Table \ref{tab:results}, when issues were first divided by type and then integrated with the LLMs for revision, higher success rates were achieved compared to the approach of selecting all types of issues from a single file and sending them to the LL in one prompt. This divided approach demonstrated effectiveness in handling bugs and vulnerabilities, where 100\% success rates were achieved. Although the comprehensive approach still yields respectable results, the divided approach generally provides higher accuracy, particularly for bugs and code smells. This is Shown in Figs. \ref{fig:success_rates} and \ref{fig:accuracy_rates} respectively. The difference in the success and accuracy rates between these approaches underscores the potential benefits of issue-specific targeting in automated code revision processes.

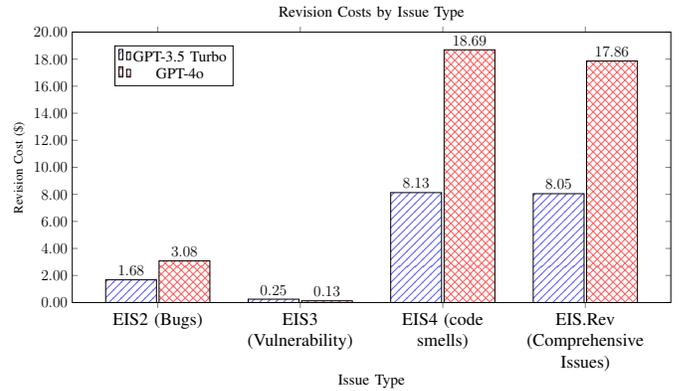
\begin{figure}[ht]
    \centering
    \resizebox{1.0\columnwidth}{!}{ 
    \begin{tikzpicture}
    \begin{axis}[
        ybar,
        bar width=40pt, 
        width=1.0\textwidth,
        height=0.5\textwidth,
        legend style={at={(0.17,0.95)}, anchor=north,font=\large},
        ylabel={Revision Cost (\$)},
        ylabel style={font=\large},
        xlabel={Issue Type},
        xlabel style={at={(0.5,-0.25)}, font=\Large},
        symbolic x coords={EIS2 (Bugs), EIS3 (Vulnerability), EIS4 (code smells), EIS.Rev (Comprehensive Issues)},
        xtick=data,
        nodes near coords,
        nodes near coords align={vertical},
        every node near coord/.append style={font=\large}, 
        ymin=0, ymax=20,
        title={Revision Costs by Issue Type},
        title style={font=\large},
        tick label style={font=\large},
        label style={font=\large},
        x tick label style={text width=3.5cm, align=center, font=\Large}, 
        enlarge x limits=0.2, 
        ylabel near ticks, 
        y tick label style={/pgf/number format/.cd, fixed, fixed zerofill, precision=2, /tikz/.cd}, 
    ]
    \addplot[pattern=north east lines, pattern color=blue!60] coordinates {(EIS2 (Bugs), 1.68) (EIS3 (Vulnerability), 0.25) (EIS4 (code smells), 8.13) (EIS.Rev (Comprehensive Issues), 8.05)};
    \addplot[pattern=crosshatch, pattern color=red!60] coordinates {(EIS2 (Bugs), 3.08) (EIS3 (Vulnerability), 0.13) (EIS4 (code smells), 18.69) (EIS.Rev (Comprehensive Issues), 17.86)};
    \legend{GPT-3.5 Turbo, GPT-4o}
    \end{axis}
    \end{tikzpicture}}
    \caption{Revision Costs by Issue Type for GPT-3.5 Turbo and GPT-4o}
    \label{fig:revision_costs}
\end{figure}

As shown in Fig. \ref{fig:revision_costs}, the cost implications of using these models for code revision were analyzed for both divided and comprehensive approaches. In the divided approach, for bugs (EIS2), a total cost of \$4.76 was incurred, with GPT-4o \$3.08 being more expensive than GPT-3.5 Turbo \$1.68. The lowest total cost was observed for vulnerabilities (EIS3) at \$0.38, with GPT-3.5 Turbo \$0.25 slightly more expensive than GPT-4o \$0.13. The highest total cost is associated with code smells (EIS4) at \$26.82, with GPT-4o \$18.69 significantly more expensive than GPT-3.5 Turbo \$8.13. The total cost of the divided approach across all issue types was to \$31.96.

In contrast, the comprehensive approach (EIS.Revised) resulted in a total cost of \$25.91, with GPT-4o accounting for \$17.86 and GPT-3.5 Turbo for \$8.05. This approach proved to be more cost-effective overall, saving approximately 18.9\% compared with the divided approach. However, it is important to note that this cost savings should be weighed against the potentially lower accuracy and success rates observed in the comprehensive approach, as discussed earlier. 

\begin{figure}[ht]
    \centering
    \includegraphics[width=1.0\linewidth]{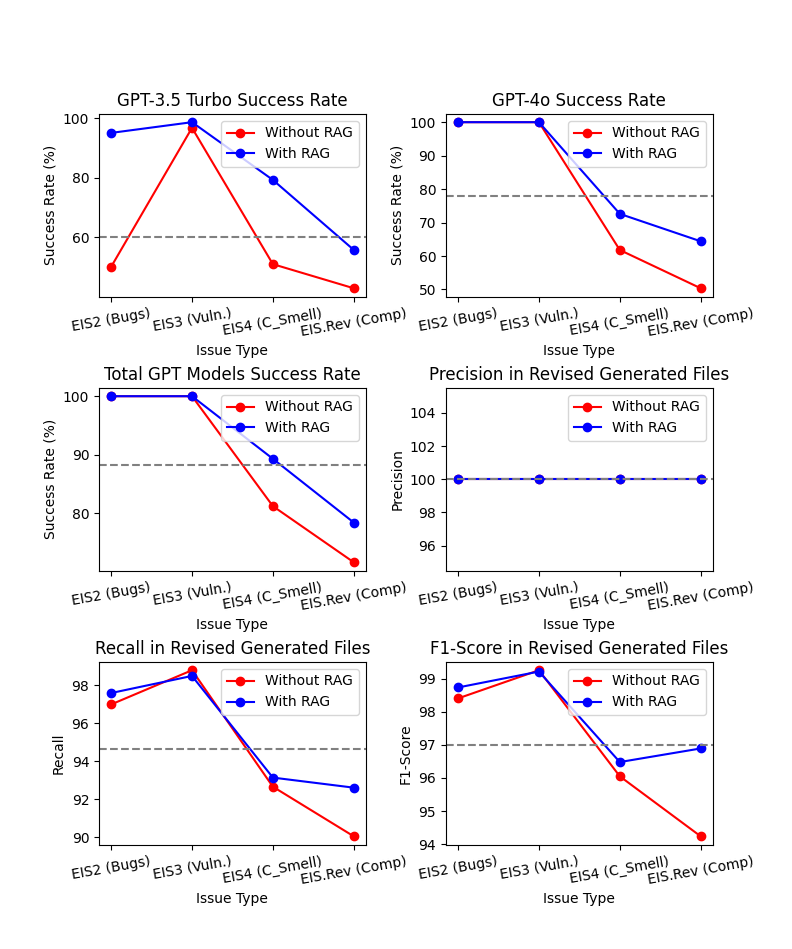}
    \caption{Comparison of success rates, precision, recall, and F1-scores for GPT-3.5 Turbo and GPT-4o with and without Retrieval-Augmented Generation (RAG) across different issue types}
    \label{fig:RAG_results}
\end{figure}

As all charts reflect the approach without applying the RAG technique, improved results were obtained after incorporating the RAG and rerunning the experiment. These results clarify the benefits of RAG in enhancing code revision. The improved outcomes and detailed results are shown in Fig. \ref{fig:RAG_results}. As shown, the number of files that the cost-effective LLM (GPT-3.5 Turbo) could be increased compared to the non-RAG approach, and also improved the success rate for GPT-4o. Consequently, using RAG in this experiment enhanced the results of the evaluation metrics.

In summary, the results demonstrate the effectiveness of the GPT models in addressing various code issues, with GPT-4o outperforming GPT-3.5 Turbo across all categories. The divided approach to issue resolution is significantly more efficient than the comprehensive method, particularly for complex issues such as code smells. However, this
efficiency comes at a higher cost, particularly when handling extensive code smells. The high F1-Score of the revised files confirms that these models, when used in a divided approach, are reliable tools for automated code revision, providing substantial time and resource savings in software development.

Although the comprehensive approach is less accurate than the divided method, it can be advantageous in scenarios where cost-efficiency is more important than precision. By addressing all the issues in a single prompt, the comprehensive method simplifies the process, making it an attractive option for users to prioritize efficiency. Moreover, integrating retrieval-augmented generation (RAG) to enhance prompts and incorporate solutions from external sources significantly increases the number of files revised and improves the F1-Score, leading to more accurate and effective revisions. 

\section{Limitations and Future Work}

This study had several limitations that highlight opportunities for future research and improvement. 

First, experiments were conducted exclusively using OpenAI's GPT models. Future research should include other models, such as Google's Gemini, Anthropic's Claude, and various versions of open-source LLaMA models. A comprehensive comparative analysis of these models will help to identify the most effective options for the application of automated code revision.

Another critical area for improvement is the enhancement of the Retrieval-Augmented Generation (RAG) technique to optimize LLM responses. Future studies should focus on refining prompt engineering techniques, including role-playing and guided responses, to generate more accurate and contextually relevant outputs. Additionally, fine-tuning models on unresolved issues could further tailor their responses to specific contexts, thereby increasing the precision and utility of the generated revisions. Integrating LLMs into CI/CD (Continuous Integration / Continuous Deployment) pipelines provides real-time feedback, enabling automated corrections and improving overall code quality. Furthermore, combining LLM generated revisions with automated testing frameworks can enhance the reliability of the code and prevent new issues from emerging.

A notable limitation of this study was the lack of test cases for automated verification of the files generated by LLMs. Addressing this issue in future work should involve generating test cases for the original files using LLMs as the initial step. These test cases can then be used to verify the original files and subsequently validate the LLM-generated revisions. By incorporating automated test case generation, the manual verification step illustrated in Fig. \ref{pipeline} can be eliminated, reducing human intervention to a single step to address residual issues or errors after LLM revisions. This streamlined approach significantly enhances the efficiency and scalability of the pipeline.

\section*{Conclusion}

This paper demonstrates the transformative potential of Large Language Models (LLMs) for automated code revision. GPT-3.5 Turbo and GPT-4o effectively resolved a wide range of code issues using prompt engineering and Retrieval-Augmented Generation (RAG) techniques, with the divided approach proving superior in both effectiveness and cost-efficiency. Although GPT-4o outperformed all categories, it came at a higher cost. Notably, this experiment shows that using LLMs to revise code issues can lead to significant time and cost savings in the software quality check process; for instance, revising over 7,500 issues took less than three hours and cost less than \$35. Future research should explore diverse LLMs and advanced design patterns to specialize in LLMs. Integrating LLM-based systems with CI/CD pipelines offers promising opportunities for real-time code improvement. Overall, LLMs show immense potential in revolutionizing code integrity assurance and promise significant time and resource savings in development processes. As these technologies advance, they become indispensable tools in maintaining high software quality standards.

\section*{Acknowledgment}  
The authors would like to thank Linda Cato and Bruce Wilkins from Team Eagle for their valuable feedback throughout this work. Additionally, the authors acknowledge the support of the SmartDelta project\footnote{\url{https://smartdelta.org}}.

\newpage
\bibliography{Reference}
\bibliographystyle{IEEEtran}
\end{document}